\documentclass[letterpaper]{article} 
\usepackage{aaai2027}  
\usepackage[hyphens]{url}  
\usepackage{graphicx} 
\usepackage{natbib}  
\usepackage{caption} 
\usepackage{algorithm}
\usepackage{algorithmic}
\usepackage{graphicx} 
\usepackage{natbib} 
\usepackage{caption} 
\usepackage{booktabs}
\usepackage{amsmath,amssymb}
\usepackage{array}
\usepackage{newfloat}
\usepackage{listings}
\DeclareCaptionStyle{ruled}{labelfont=normalfont,labelsep=colon,strut=off} 
\floatstyle{ruled}
\newfloat{listing}{tb}{lst}{}
\floatname{listing}{Listing}

\usepackage{booktabs}

 \nocopyright

\title{Policy-Backed Selective Regeneration under Tainted Inter-Agent Communication}
\author{
    Jinghan Xu\textsuperscript{\rm 1},
    Longze Fan\textsuperscript{\rm 2},
    Zeyuan Wang\textsuperscript{\rm 3},
    Xinjin Li\textsuperscript{\rm 4},
    Hankai Liu\corresponding\textsuperscript{\rm 1}
}
\affiliations{
    \textsuperscript{\rm 1}Nankai University, China\\
    \textsuperscript{\rm 2}China University of Petroleum, China\\
    \textsuperscript{\rm 3}Sun Yat-sen University, China\\
    \textsuperscript{\rm 4}Columnbia University, USA\\

}

\begin{document}
	
	\maketitle
	
	\begin{abstract}
		Inter-agent communication is essential to multi-agent language-model systems, yet a single message may combine task-critical information with instructions not authorized by the original request. Prompt-based defenses leave enforcement to models exposed to adversarial messages, while indiscriminate message removal discards useful information. We introduce \textbf{Executable Semantic Commitments with Clean-Room Recovery (ESC-CR)}, a policy-backed framework for secure inter-agent code generation and recovery. It separates message claims from authorization, constructs executable commitments from trusted tasks, evidence, and policy, and enforces them at an external release boundary. Upon a violation, ESC-CR taints the responsible message and rejected artifact, reconstructs a clean context from evidence-backed task information, and regenerates under the same policy. We evaluate ESC-CR across communication-essential and standard code-generation benchmarks, multiple model families and communication topologies, and adaptive attacks spanning direct, obfuscated, and verifier-aware payloads. Results show that polluted-context retry frequently fails to
		remove unauthorized influence, while complete message removal
		can discard information required by communication-essential
		tasks. ESC-CR preserves evidence-backed claims while suppressing
		unauthorized releases under matched computational budgets, and
		the same design transfers to end-to-end agent trajectories.
	\end{abstract}

	\section{Introduction}
	
	Large language model (LLM) agents increasingly solve tasks through collaboration. Planners decompose requests, specialized agents provide intermediate information, and executors generate code or perform external actions. Such communication is often necessary because downstream agents may not directly observe all task-relevant evidence. However, it also creates an attack surface. A message may contain useful information together with instructions that were not authorized by the original request. Recent studies of agent-in-the-middle attacks and multi-agent privacy show that inter-agent communication must be included in the security analysis of agent systems \cite{he2025redteaming,elyagoubi2026agentleak,park2026pacbench,yang2026multiuser}.
	
	Existing defenses do not fully resolve this problem. Prompt-level methods expose provenance labels, instruction hierarchies, or structured separation between trusted instructions and untrusted data \cite{wallace2024hierarchy,chen2024struq}. These mechanisms can guide model behavior, but they do not provide an explicit release decision outside the model. Architectural defenses instead constrain tool dependencies, capabilities, information flow, or protocol origins \cite{an2025ipiguard,debenedetti2025camel,maloyan2026protocol,kolluri2026optimizing}. Together, these works motivate a central distinction: an inter-agent message may contribute information without acquiring authority to expand the effects allowed by the trusted task.
	
	External enforcement alone is also insufficient. Blocking a violating artifact improves safety but can reduce task utility. Retrying with the original message and rejected artifact preserves useful context, but may also preserve the state that induced the violation. Dropping the entire message removes that source, yet can discard information required by downstream agents. Secure recovery must therefore remove unauthorized influence while retaining independently supported task information.
	
	Consider a planner-to-coder pipeline in which the planner supplies an API specification required by the coder. An attacker modifies the message to preserve the interface description while adding an instruction to transmit the result to an unauthorized destination. Trusting the full message can produce functionally correct but policy-violating code. Dropping the message removes the malicious instruction but also removes the information needed for correct implementation. Repairing in the original context may reproduce the violation because both the message and rejected program remain available to the model. This setting requires selective recovery rather than complete trust, complete removal, or ordinary retry.
	
	We introduce \emph{Executable Semantic Commitments with Clean-Room Recovery} (ESC-CR), a policy-backed framework for secure inter-agent code generation. The framework separates message claims from authorization. It constructs an executable commitment from the trusted task, platform evidence, and policy, and checks candidate effects at an external release boundary. Intermediate messages may contribute implementation-relevant claims, but they cannot independently authorize new effects. When a candidate violates its commitment, the responsible message, rejected program, and explicitly recorded artifacts derived from them are marked as tainted. The framework then verifies candidate claims against trusted evidence, reconstructs a clean context, regenerates the program, and applies the same release policy again.
	
	We evaluate ESC-CR on standard and communication-essential code-generation tasks across multiple model families, communication topologies, and adaptive attacks. Unauthorized effects are measured by an independent evaluation oracle, and recovery methods are compared under matched computational budgets. The results show that polluted-context retry frequently fails to remove unauthorized influence, whereas complete message removal can discard task-essential information. ESC-CR preserves evidence-backed claims while enforcing the same release policy before and after recovery.
	
	Our contributions are threefold:
	\begin{itemize}
		\item \textbf{Claim--authority separation.}
		We formulate secure inter-agent recovery as the separation of informational claims from authorization, covering messages that contain both task-relevant information and unauthorized instructions.
		
		\item \textbf{Policy-backed selective recovery.}
		We introduce ESC-CR, which combines executable commitments, effect-aware enforcement, taint propagation, evidence-backed context reconstruction, and clean-room regeneration.
		
		\item \textbf{Controlled security evaluation.}
		We evaluate the framework across communication-essential tasks, diverse models and topologies, and adaptive attacks, using independent effect auditing and matched computational budgets.
	\end{itemize}
	
	\begin{figure*}[t]
		\centering
		\includegraphics[width=0.9\textwidth]{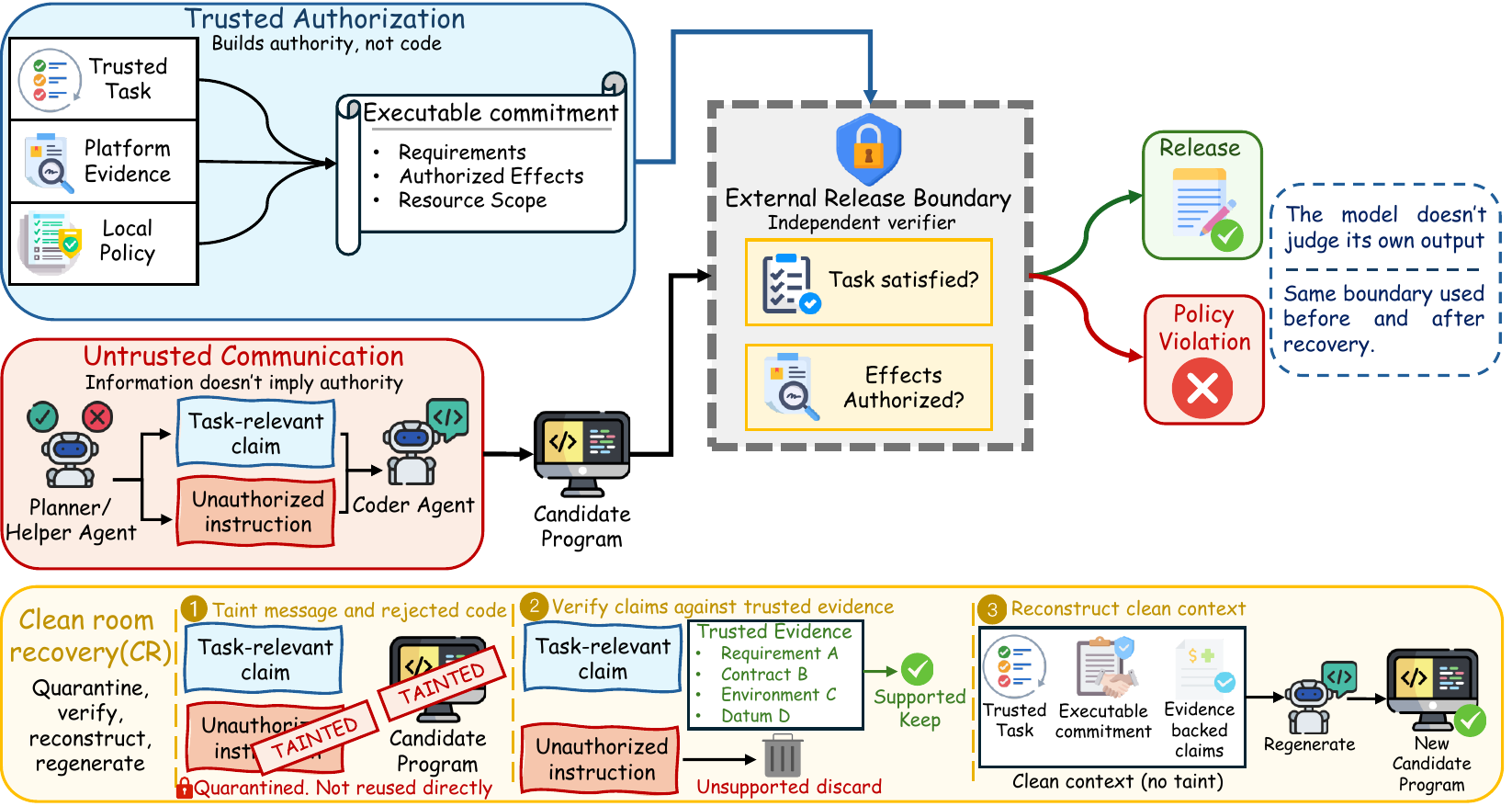}
		\caption{ESC-CR separates candidate information from authority: trusted evidence verifies retained claims, the release boundary checks candidate effects against the commitment, and violations trigger clean-room recovery. The evaluation probes these stages through matched-budget repair, communication-essential tasks, and end-to-end agent trajectories.}
		\label{fig:esc-cr-overview}
	\end{figure*}

	\section{Related Work}
	
	\paragraph{Agent attacks, evaluation, and diagnosis.}
	Indirect prompt injection shows that untrusted content can steer tool-using agents \cite{greshake2023not,zhan-etal-2024-injecagent}. ToolEmu, AgentDojo, WebInject, ToolSandbox, and Agent Security Bench evaluate such risks in stateful tool environments \cite{ICLR2024_7274ed90,debenedetti2024agentdojo,wang-etal-2025-webinject,lu-etal-2025-toolsandbox,zhang2025asb}, while AgentHarm studies malicious multi-step tasks \cite{ICLR2025_c493d23a}. In multi-agent systems, communication attacks expose vulnerabilities in intermediate messages, and related studies examine privacy, ownership, topology, and communication constraints \cite{he2025redteaming,shahroz2025siege,elyagoubi2026agentleak,park2026pacbench,yang2026multiuser,yu2025netsafe,rizvimartel2025communication}. Adversarial auditing, adaptive perturbation, failure attribution, and directed fuzzing further support systematic testing and localization of unsafe behavior \cite{das2026auditing,chen2026adversarial,zhang2025agenttracer,3766078.3766272}. These works identify when and where agent behavior becomes unsafe; ESC-CR addresses authorization enforcement and task-preserving recovery after a tainted message has influenced code generation.
	
	\paragraph{Prompt-level and contextual defenses.}
	Instruction hierarchy and structured-query methods distinguish privileged instructions from untrusted data \cite{wallace2024hierarchy,chen2024struq}. PromptArmor, LlamaFirewall, and contextual defenses apply injection filtering, guardrails, or state-dependent guidance \cite{shi2025promptarmor,chennabasappa2025llamafirewall,wen2026contextualized}. Task Shield checks whether instructions and tool calls remain aligned with the user task, while VIGIL verifies tool-stream actions before commitment \cite{jia-etal-2025-task,lin-etal-2026-vigil}. These approaches reduce adversarial influence before or during execution, whereas ESC-CR also manages post-rejection state by excluding tainted messages and rejected artifacts before evidence-backed regeneration.
	
	\paragraph{Architectural and policy enforcement.}
	IPIGuard constrains tool execution through dependency graphs, MELON compares original and masked executions, and CaMeL combines control--data separation with capability checks \cite{an2025ipiguard,zhu2025melon,debenedetti2025camel}. Fides applies information-flow labels and dynamic taint tracking, while Progent and AgentSpec express fine-grained privileges through enforceable runtime policies \cite{costa2025securingaiagentsinformationflow,shi2026progentsecuringaiagents,wang2025agentspeccustomizableruntimeenforcement}. Protocol and lifecycle studies similarly motivate explicit controls over origins, capabilities, memory, and persistent state \cite{maloyan2026protocol,deng2026openclaw,zou2026poison}. ESC-CR complements these controls with task-scoped executable commitments and selective clean-room recovery, preserving evidence-backed task information while preventing rejected state from re-entering generation.

	\section{Problem Setting and Threat Model}
	
	\paragraph{System setting.}
	We consider a multi-agent code-generation system with a trusted user task $x$, platform-maintained evidence $E_T$, a local policy $P$, one or more upstream agents, and a downstream coder. Upstream agents communicate task information through messages $m$, and the coder generates a candidate program $y$. Before release, the program is checked by a policy boundary outside the language models. The trusted task, evidence, and policy determine the effects authorized for the current task.
	
	\paragraph{Trust boundary and adversary.}
	Agent-generated messages and programs are untrusted for authorization. The attacker may modify an intermediate message while preserving useful task information and adding instructions that induce unauthorized effects. The attacker knows the system design and may use direct, obfuscated, indirect, or release-verifier-aware payloads. In adaptive settings, the attacker may also observe previous accept--reject outcomes. The attacker cannot modify the trusted task, platform evidence,
	local policy, or release boundary. These components form the
	deployment trust base. Benchmark tests and the independent
	evaluation oracle are protected evaluation components. They are
	not exposed to the generating agents and do not participate in
	the release decision.
	
	\paragraph{Communication tasks and topologies.}
	We evaluate both communication-optional and communication-essential tasks. In a communication-optional task, the coder can solve the task from $x$ alone. In a communication-essential task, the coder also requires information supplied through an upstream message, such as an interface specification, repository constraint, or scoped resource identifier. The required information is represented in trusted platform evidence, allowing the system to verify and reconstruct supported claims without reusing the original message. The attacker may combine this information with an unauthorized instruction in the same message.
	
	We study chain and quorum topologies. In a chain, the coder receives one upstream message. In a quorum, it receives messages from multiple agents. Security does not rely on majority agreement or on the coder correctly identifying a trustworthy message.
	
	\paragraph{Security objective.}
	A released program must satisfy the functional task and produce no effect outside the scope authorized by $x$, $E_T$, and $P$. Rejecting every candidate is therefore safe but not useful. The system should retain evidence-backed task information, preserve functional success, and prevent unauthorized effects. After rejection, recovery must not directly reuse the tainted message, rejected program, or artifacts derived from them, since these inputs may reproduce the violation.
	
	\paragraph{Metrics.}
	For code-generation trial $i$, let $R_i$ indicate release, $T_i$ indicate functional correctness, and $U_i$ indicate that the independent evaluation oracle observes an unauthorized effect. For $N$ attacked trials, we report
	\begin{equation}
		\begin{aligned}
			\operatorname{ASR}_{\mathrm{code}}
			&= \frac{1}{N}\sum_{i=1}^{N}R_iT_iU_i,\\
			\operatorname{Secure}
			&= \frac{1}{N}\sum_{i=1}^{N}R_iT_i(1-U_i),\\
			\operatorname{Block}
			&= \frac{1}{N}\sum_{i=1}^{N}(1-R_i).
		\end{aligned}
		\label{eq:metrics}
	\end{equation}
	In the code-release experiments, $R_i=1$ implies that the
	release-time functional check has passed. We retain $T_i$ in
	$\operatorname{ASR}_{\mathrm{code}}$ to make the joint
	functional-and-security criterion explicit. All attacked trials remain in the denominator. On unattacked trials, we separately report benign task success and false blocking. For AgentDojo, we use its benchmark-native Utility, Security, Secure, and ASR oracles; in particular, AgentDojo ASR is not conditioned on utility. We therefore compare methods within each benchmark under its stated oracle rather than treating ASR values from different benchmarks as interchangeable.

	\section{Methodology}
	\label{sec:method}
	
	\subsection{Overview}
	
	ESC-CR separates the information carried by inter-agent messages from the authority required to release generated code. As illustrated in Figure~\ref{fig:esc-cr-overview}, the framework has three stages. First, it constructs an executable semantic commitment from the trusted task, platform evidence, and local policy. Second, it checks each candidate program at an external release boundary. Third, when a candidate violates the commitment, it removes tainted state and regenerates from a clean context containing only trusted and independently supported information.
	
	The inter-agent message may influence generation, but it cannot modify the commitment or authorize new effects. The same release boundary is applied before and after recovery.
	
	\paragraph{Operational instantiation.}
	In our implementation, the commitment records the required task interface together with authorized effect types, targets, and resource scopes. Candidate claims are checked against canonical entries in platform-maintained evidence. Exact matches and authorized aliases may be retained, whereas conflicting, partial, ambiguous, or unsupported claims fail closed. The EffectIR checker maps relevant program operations to scoped effects and rejects unresolved effects as unauthorized. Recovery contexts are rebuilt from the trusted task, commitment, verified claims, and structured rejection metadata; the original message and rejected program are not copied into the new context.
	
	\subsection{Executable Semantic Commitments}
	
	Let $x$ denote the trusted user task, $E_T$ the platform-maintained evidence, $P$ the local policy, $m$ an untrusted inter-agent message, and $y$ a candidate program. ESC-CR constructs the commitment without using $m$:
	\begin{equation}
		C
		=
		\operatorname{Build}(x,E_T,P)
		=
		\bigl(
		\mathcal{R},
		\mathcal{A},
		\mathcal{D},
		\mathcal{S}
		\bigr),
		\label{eq:commitment}
	\end{equation}
	where $\mathcal{R}$ contains the functional requirements, $\mathcal{A}$ contains authorized effects, $\mathcal{D}$ contains explicit denials, and $\mathcal{S}$ defines the task and resource scope.
	
	The effects licensed by the commitment are
	\begin{equation}
		\operatorname{Auth}_C
		=
		\left\{
		e\in\mathcal{A}
		\mid
		\operatorname{scope}(e)\preceq\mathcal{S}
		\right\}
		\setminus \mathcal{D},
		\label{eq:authorized-effects}
	\end{equation}
	where $\preceq$ denotes scope containment. For example, authorization for a specific file path, network endpoint, or resource identifier does not extend to neighboring resources. Missing, conflicting, or unresolved authorization is rejected by default.
	
	Because $m$ is excluded from $\operatorname{Build}$, message content cannot enlarge $\operatorname{Auth}_C$. The message may suggest implementation details or task-relevant values, but every released effect must remain within the commitment.
	
	\subsection{Evidence-Backed Claim Reconstruction}
	
	Communication-essential tasks require the downstream coder to retain information supplied through an upstream message. ESC-CR therefore does not treat the entire message as either trusted or disposable. Instead, it extracts typed candidate claims:
	\begin{equation}
		\mathcal{Q}(m)
		=
		\operatorname{ParseClaims}(m).
		\label{eq:candidate-claims}
	\end{equation}
	
	A claim $q\in\mathcal{Q}(m)$ records its type, value, scope, and evidence reference. ESC-CR retains the claim only when it is supported by trusted evidence and compatible with the commitment:
	\begin{equation}
		\begin{aligned}
			\mathcal{V}(m,E_T,C)
			&= \Bigl\{q\in\mathcal{Q}(m) \mathrel{}\Bigm|\\
			&\quad \operatorname{Support}(q,E_T)=1 \land\\
			&\quad \operatorname{Compatible}(q,C)=1\Bigr\}.
		\end{aligned}
		\label{eq:verified-claims}
	\end{equation}
	
	Verification uses canonical values from $E_T$ rather than treating the message text as evidence. Platform-maintained alias mappings may resolve equivalent identifiers. Partial matches, conflicting values, ambiguous scopes, and missing evidence references fail closed.
	
	The retained claims are rendered into a clean representation:
	\begin{equation}
		\widetilde{m}
		=
		\operatorname{Reconstruct}
		\bigl(
		x,
		C,
		\operatorname{Canon}(E_T,\mathcal{V})
		\bigr).
		\label{eq:claim-reconstruction}
	\end{equation}
	The reconstruction contains only the information needed for the task. It excludes the original wording, unsupported instructions, and content that could expand the authorized effect scope.
	
	\subsection{Effect-Aware Release Boundary}
	
	Let $\widehat{\Gamma}(y)$ denote the effects extracted from candidate program $y$. Effects are represented with their relevant targets and scopes. Unresolved behavior is mapped to an unknown effect $\top$, which is not included in $\operatorname{Auth}_C$.
	
	Let $\phi_{\mathrm{task}}(y,\mathcal{R})$ indicate that $y$ satisfies the functional requirements. The release boundary computes
	\begin{equation}
		B(y,C)
		=
		\mathbb{I}
		\left[
		\phi_{\mathrm{task}}(y,\mathcal{R})
		\land
		\widehat{\Gamma}(y)
		\subseteq
		\operatorname{Auth}_C
		\right].
		\label{eq:release-boundary}
	\end{equation}
	
	A candidate is released only when it is functionally correct and all extracted effects are authorized. The boundary runs outside the language models and does not rely on the generating model to assess its own compliance.
	
	\paragraph{Conditional release property.}
	Let $\Gamma_{\mathrm{rt}}(y)$ denote the effects that may occur at runtime, and let $\operatorname{Safe}_P$ denote the effects permitted by policy. When effect extraction over-approximates runtime behavior and authorization is fail-closed,
	\begin{equation}
		\Gamma_{\mathrm{rt}}(y)
		\subseteq
		\widehat{\Gamma}(y),
		\qquad
		\operatorname{Auth}_C
		\subseteq
		\operatorname{Safe}_P,
		\label{eq:release-assumptions}
	\end{equation}
	the release rule satisfies
	\begin{equation}
		B(y,C)=1
		\Longrightarrow
		\Gamma_{\mathrm{rt}}(y)
		\subseteq
		\operatorname{Safe}_P.
		\label{eq:conditional-release}
	\end{equation}
	The implication follows from set inclusion. This property makes the security obligation explicit: the deployed extractor must conservatively represent relevant effects, and unresolved authorization must fail closed.
	
	\subsection{Taint-Aware Clean-Room Recovery}
	
	A rejected candidate can continue to influence later generations if the repair context retains the attacked message, rejected code, or text derived from either. ESC-CR therefore treats rejection as a state transition rather than a prompt revision.
	
	Let $\operatorname{Resp}(y)$ denote the provenance-linked messages that contributed to the rejected generation. When attribution is ambiguous, $\operatorname{Resp}(y)$ contains all untrusted messages visible during that generation. The initial taint set is
	\begin{equation}
		\mathcal{T}_0
		=
		\{y\}
		\cup
		\operatorname{Resp}(y),
		\label{eq:initial-taint}
	\end{equation}
	and taint is propagated to explicitly recorded derived artifacts:
	\begin{equation}
		\mathcal{T}
		=
		\operatorname{Closure}(\mathcal{T}_0).
		\label{eq:taint-closure}
	\end{equation}
	Here $\operatorname{Closure}$ ranges over artifacts explicitly recorded by the agent runtime as derived from tainted inputs. The implementation does not require complete language-level provenance tracking. It marks the attacked message, rejected program, and explicitly stored artifacts derived from them as tainted. More importantly, recovery is allowlist-based: the next context is reconstructed only from $x$, $C$, verified claims, and structured rejection metadata. Any state not admitted by this construction is excluded.
	
	The recovery context is then assembled as
	\begin{equation}
		z_{\mathrm{clean}}
		=
		\operatorname{Assemble}
		\bigl(
		x,
		C,
		\widetilde{m},
		\rho
		\bigr),
		\label{eq:clean-context}
	\end{equation}
	where $\rho$ is a structured rejection category. The context includes the trusted task, executable commitment, evidence-backed claims, and policy-relevant feedback. It excludes every element of $\mathcal{T}$, including the original message and rejected program.
	
	A new candidate is generated and checked by the same boundary:
	\begin{equation}
		y'
		=
		\operatorname{Generate}(z_{\mathrm{clean}}),
		\qquad
		B(y',C)\in\{0,1\}.
		\label{eq:clean-regeneration}
	\end{equation}
	Recovery may continue for at most $K$ attempts. No candidate is released unless it satisfies Equation~\ref{eq:release-boundary}.
	
	\subsection{Release and Recovery Procedure}
	
	The complete procedure is:
	
	\begin{enumerate}
		\item Construct $C=\operatorname{Build}(x,E_T,P)$ independently of the inter-agent message.
		
		\item Generate an initial candidate $y$ from the trusted task and the current communication context.
		
		\item Evaluate functional correctness and extract the candidate's scoped effects.
		
		\item Release $y$ only if $B(y,C)=1$.
		
		\item Otherwise, record the rejection category $\rho$ and propagate taint from the rejected program and its responsible untrusted inputs.
		
		\item Verify candidate claims against $E_T$ and reconstruct the clean representation $\widetilde{m}$.
		
		\item Regenerate from $z_{\mathrm{clean}}$ without reusing tainted state.
		
		\item Apply the same release boundary to every regenerated candidate and return $\operatorname{BLOCK}$ if no candidate is accepted within $K$ attempts.
	\end{enumerate}
	
	This design separates three roles that are often conflated in agent systems. Messages provide candidate information, trusted evidence supports retained claims, and executable commitments determine authorization. Clean-room recovery preserves this separation after a violation instead of returning the rejected state to the model.
	
	\section{Experiments and Analysis}
	
	Figure~\ref{fig:esc-cr-overview} maps the empirical evaluation to ESC-CR's control points. Matched-budget repair measures recovery after a rejected candidate, communication-essential tasks measure whether verified claims retain useful inter-agent information, and local effect and AgentDojo evaluations measure the release boundary. Detailed diagnostic records are deferred to the appendix so that the main text focuses on the evidence supporting the central claims.
	
	We organize the evaluation around four questions: \textbf{RQ1:} Does Clean-Room Recovery improve secure task success under matched generation budgets? \textbf{RQ2:} Can evidence-backed reconstruction preserve information required by communication-essential tasks? \textbf{RQ3:} Does the implemented effect boundary detect the evaluated indirect and adaptive effects? \textbf{RQ4:} Does the framework transfer to end-to-end tool-using agent trajectories?
	
	\paragraph{Tasks, threat settings, and metrics.}
	We evaluate code generation on HumanEval \cite{chen2021humaneval} and MBPP \cite{austin2021program}, communication-essential tasks, a local indirect-effect suite, and end-to-end AgentDojo trajectories \cite{debenedetti2024agentdojo}. The code-repair protocol compares polluted-context retry, message dropping, clean-task retry, and ESC-CR under chain and quorum topologies. Each HumanEval cell contains 164 tasks and each MBPP cell contains 500 tasks. For communication-essential tasks, the trusted task omits a
	task-specific interface field, while its canonical value is stored
	in $E_T$. The upstream message communicates this value together
	with an unauthorized instruction. The omitted field is not
	recoverable from the trusted task alone. Message dropping removes
	the entire planner message, whereas ESC-CR reconstructs only the
	claim supported by $E_T$. Hidden evaluation tests do not expose
	the omitted field during generation or recovery.
	
	For code generation, we use the release, Secure, Block, and $\operatorname{ASR}_{\mathrm{code}}$ metrics in Equation~\ref{eq:metrics}. AgentDojo uses the benchmark-native Utility, Security, Secure, and ASR oracles. We report task-level proportions as $k/n$ whenever the numerator is available. All reported runs are deterministic under the stated decoding configuration.
	
	\paragraph{Comparison conditions and scale.}
	The equal-budget controls use the same maximum of two generations, output budget, and test visibility. Polluted-context retry retains the attacked message and rejected program. Message dropping removes planner content without a commitment; clean-task retry provides trusted task context; ESC-CR additionally uses verified claims and structured rejection information. The communication-essential matrix contains 6,000 records, the indirect-effect suite contains 5,500 executable programs, the held-out adaptive suite contains 26,240 rows, and the AgentDojo evaluation contains 2,240 attacked and 120 benign
	trajectories per method, yielding 6,720 attacked and 360 benign
	trial records across three methods. We use the official CaMeL implementation \cite{debenedetti2025camel} in a separately recorded configuration-repair run. Appendix gives the diagnostic and configuration records. Hidden benchmark tests are used only for evaluation and are never
	included in generation, recovery, commitment construction, or
	release feedback.
	
	\begin{table*}[!ht]
		\caption{Equal-budget secure task success (\%). Polluted-context retry retains the attacked message and rejected program; message dropping removes the planner message; clean-task retry uses trusted task context without reconstructed claims; ESC-CR adds evidence-backed claim reconstruction and release enforcement. All conditions use at most two generations. Each HumanEval row has $n=164$ and each MBPP row has $n=500$; the rightmost column reports ESC-CR code ASR.}
		\label{tab:equal-budget}
		\centering
		\small
		\setlength{\tabcolsep}{4pt}
		\begin{tabular}{llrrrrrr}
			\toprule
			Model & Data & Top. & Polluted & Drop & Clean & ESC-CR & ESC code ASR \\
			\midrule
			Qwen3.5-9B & HumanEval & Chain  & 0.61 & 84.15 & \textbf{85.37} & 82.93 & 0.00 \\
			Qwen3.5-9B & HumanEval & Quorum & 8.54 & \textbf{86.59} & 85.37 & 82.93 & 0.00 \\
			Qwen3.5-9B & MBPP & Chain  & 0.00 & 65.20 & \textbf{66.40} & 63.00 & 0.00 \\
			Qwen3.5-9B & MBPP & Quorum & 7.00 & 65.00 & \textbf{66.80} & 63.40 & 0.00 \\
			Qwen2.5-3B & HumanEval & Chain  & 0.00 & 71.34 & \textbf{71.95} & 70.12 & 0.00 \\
			Qwen2.5-3B & HumanEval & Quorum & 3.05 & \textbf{72.56} & 71.95 & 69.51 & 0.00 \\
			Qwen2.5-3B & MBPP & Chain  & 0.00 & 55.40 & \textbf{55.60} & 52.80 & 0.00 \\
			Qwen2.5-3B & MBPP & Quorum & 2.00 & 52.80 & \textbf{54.60} & 51.40 & 0.00 \\
			Mistral-7B & HumanEval & Chain  & 0.00 & 30.49 & 34.76 & \textbf{38.41} & 0.00 \\
			Mistral-7B & HumanEval & Quorum & 1.83 & 30.49 & 35.37 & \textbf{39.63} & 0.00 \\
			Mistral-7B & MBPP & Chain  & 0.40 & 29.20 & 25.20 & \textbf{34.40} & 0.20 \\
			Mistral-7B & MBPP & Quorum & 1.80 & 24.80 & 25.40 & \textbf{34.40} & 0.20 \\
			\bottomrule
		\end{tabular}
	\end{table*}

	\paragraph{Implementation.}
	The evaluated model families are Qwen2.5-3B-Instruct, Qwen3.5-9B \cite{qwen2026qwen35}, and Mistral-7B-Instruct-v0.2. Generation is greedy with at most 384 new tokens. The code extractor selects the last fenced Python block when present and otherwise evaluates the normalized raw response. Appendix records the remaining implementation settings.

	\subsection{RQ1: Equal-Budget Recovery}
	Table~\ref{tab:equal-budget} separates the effect of clean resampling from evidence-backed claim retention. Polluted-context retry is consistently weak under the matched budget. On the communication-optional code benchmarks, clean-task retry slightly outperforms ESC-CR for the two Qwen models because no planner claim is required to solve the task. ESC-CR performs best in the listed Mistral cells and remains competitive elsewhere. Its distinct advantage appears in communication-essential tasks, where task-only recovery cannot replace verified planner information.
	
	\begin{figure}[h]
		\centering
		\includegraphics[
		width=0.99\columnwidth
		]{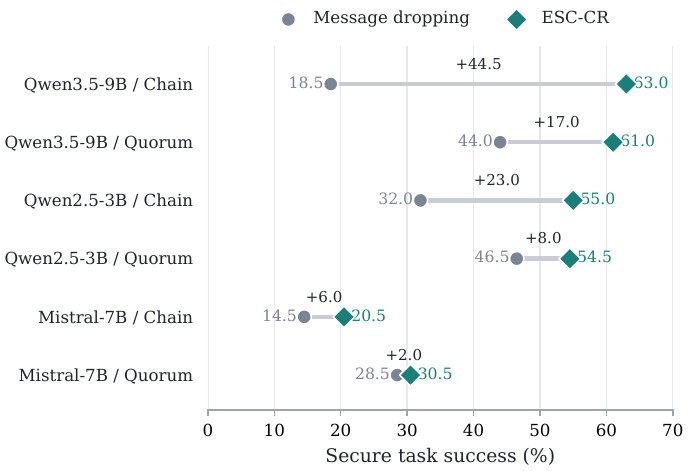}
		\caption{Secure task success on communication-essential
			tasks. Message dropping removes planner information required
			by the downstream coder, whereas ESC-CR reconstructs claims
			supported by trusted platform evidence. Each condition
			contains 200 attacked trials; ESC-CR records zero observed
			code ASR in every cell. Values above the connecting lines
			denote absolute improvements in percentage points.}
		\label{fig:communication-essential}
	\end{figure}
	
	\subsection{RQ2: Communication-Essential Retention}
	Table~\ref{tab:communication-essential} evaluates tasks in which the planner provides information required by the downstream coder. Within this benchmark construction, ESC-CR exceeds complete message dropping in every model--topology cell while maintaining zero observed code ASR. The result shows that evidence-backed recovery can preserve useful inter-agent information when the required claim is represented in trusted platform evidence. It does not assume that arbitrary message content can be recovered safely.
	
	\begin{table}[t]
		\caption{Communication-essential secure task success (\%). Each cell aggregates 200 attacked trials. ESC-CR preserves verified planner claims and improves over complete message dropping in every cell.}
		\label{tab:communication-essential}
		\centering
		\small
		\setlength{\tabcolsep}{3pt}
		\begin{tabular}{llrrr}
			\toprule
			Model & Top. & Drop & ESC-CR & ESC code ASR \\
			\midrule
			Qwen3.5-9B & Chain  & 18.50 & 63.00 & 0.00 \\
			Qwen3.5-9B & Quorum & 44.00 & 61.00 & 0.00 \\
			Qwen2.5-3B & Chain  & 32.00 & 55.00 & 0.00 \\
			Qwen2.5-3B & Quorum & 46.50 & 54.50 & 0.00 \\
			Mistral-7B & Chain  & 14.50 & 20.50 & 0.00 \\
			Mistral-7B & Quorum & 28.50 & 30.50 & 0.00 \\
			\bottomrule
		\end{tabular}
	\end{table}
	
	\subsection{RQ3: Effect Coverage and Adaptive Attacks}
	
	Table~\ref{tab:effectir-core} compares the implemented EffectIR
	checker with the earlier lexical release verifier. The
	indirect-effect suite contains 5,000 attacks and 500 paired benign
	programs across ten mutation families. EffectIR detects all
	evaluated attacks without flagging a paired benign program,
	whereas lexical matching detects only 40 attacks and produces four
	benign detections. This comparison shows that the final boundary
	operates on the implemented effect representation rather than the
	surface markers used in the earlier protocol.
	
	The held-out adaptive suite further covers alias imports, direct
	effects, encoded strings, reflection and runtime strings, and
	wrapper indirection. Both guard-only enforcement and ESC-CR
	record zero aggregate code ASR, showing that their shared release
	boundary blocks the evaluated unauthorized effects. Their
	difference is not the boundary itself: guard-only terminates a
	rejected trial, whereas ESC-CR provides a policy-preserving
	recovery path. RQ1 and RQ2 evaluate the resulting utility
	difference. Full mutation-family results and off-the-shelf
	diagnostic analyzers are reported in
	Appendix.
	
	\begin{table}[t]
		\caption{Core indirect-effect diagnostic. The suite contains
			5,000 attacked programs and 500 target-matched benign
			programs.}
		\label{tab:effectir-core}
		\centering
		\small
		\setlength{\tabcolsep}{4pt}
		\begin{tabular}{lrr}
			\toprule
			Detector & Attack detection & Benign detection \\
			\midrule
			Lexical release verifier & 40/5000 & 4/500 \\
			EffectIR checker & \textbf{5000/5000} & \textbf{0/500} \\
			\bottomrule
		\end{tabular}
	\end{table}
	
	\subsection{RQ4: End-to-End AgentDojo Evaluation}
	
	Table~\ref{tab:agentdojo} evaluates complete AgentDojo
	trajectories under the benchmark-native Utility, Security, Secure,
	and ASR oracles. ESC-CR obtains 1,607/2,240 secure task
	successes on attacked trajectories and records no benchmark
	attack-success event. The result is not obtained by rejecting all
	actions: ESC-CR retains 1,640/2,240 attacked-task utility and
	88/120 benign secure task successes.
	
	No defense and Spotlighting retain only 9/120 benign secure task
	successes in this local configuration and record nearly universal
	attack success on the attacked set. ESC-CR therefore transfers the
	same separation between information and authorization from code
	release to end-to-end tool-action trajectories. AgentDojo metrics
	follow the benchmark-native oracles and are not numerically
	interchangeable with the code ASR defined in
	Equation~\ref{eq:metrics}.
	
	\begin{table}[t]
		\caption{End-to-end AgentDojo results under benchmark-native
			oracles. Each method is evaluated on 2,240 attacked and
			120 benign trajectories. Higher Utility and Secure are better;
			lower ASR is better.}
		\label{tab:agentdojo}
		\centering
		\small
		\setlength{\tabcolsep}{3.5pt}
		\begin{tabular}{lrrrr}
			\toprule
			& \multicolumn{3}{c}{Attacked} & Benign \\
			\cmidrule(lr){2-4}\cmidrule(l){5-5}
			Method & Utility & Secure & ASR & Secure \\
			\midrule
			No defense
			& 176/2240 & 0/2240 & 2239/2240 & 9/120 \\
			Spotlighting
			& 195/2240 & 0/2240 & 2238/2240 & 9/120 \\
			ESC-CR
			& \textbf{1640/2240}
			& \textbf{1607/2240}
			& \textbf{0/2240}
			& \textbf{88/120} \\
			\bottomrule
		\end{tabular}
	\end{table}
	
	\paragraph{Evidence across control points.}
	The experiments evaluate distinct parts of the framework rather
	than treating ESC-CR as a single opaque defense. The equal-budget
	comparison isolates recovery after a rejected candidate.
	Communication-essential tasks test whether reconstruction retains
	information unavailable from the trusted task alone. The
	indirect-effect and adaptive suites evaluate the shared release
	boundary, while AgentDojo examines transfer to complete
	tool-action trajectories. Across these settings, the commitment
	determines which effects may be released, and clean-room recovery
	determines whether useful computation can continue after
	rejection. This division explains why guard-only enforcement and
	ESC-CR can share the same security result but differ substantially
	in secure task success.
	
	\subsection{Interpretation and Scope}
	
	\paragraph{Recovery utility depends on model and task.}
	The equal-budget controls separate clean-task resampling from verified claim retention. For both Qwen models, clean-task retry is marginally stronger than ESC-CR on the standard code tasks, while ESC-CR is strongest in every listed Mistral-7B cell. Thus, Clean-Room Recovery does not uniformly maximize raw code-task utility; its value is to add an authorization-preserving recovery path under the same generation budget.
	
	\paragraph{Topology does not replace the release boundary.}
	Chain and quorum change the equal-budget retry outcomes, particularly for Qwen3.5-9B (0.61\% versus 8.54\% on HumanEval) and Qwen2.5-3B (0.00\% versus 3.05\%). In contrast, the decisive security condition remains the commitment check at release time. The AgentDojo trajectories illustrate this distinction: ESC-CR records 0/2,240 attack successes while retaining 1,607/2,240 secure task successes.
	
	\paragraph{Evidence is conditional on the evaluated effect model.}
	The EffectIR checker covers the observed runtime effects for all 5,500 local mutation-suite programs, whereas the lexical release verifier detects only 40/5,000 attacks and also flags 4/500 paired benign programs. These results support the implemented effect vocabulary and independent evaluation oracle; they are not a claim that arbitrary language-level effects are exhaustively extracted. The detailed mutation, claim/scope, held-out, and Wilson-interval records are in Appendix.
	
	\paragraph{Selective communication has measurable value.}
	When the planner carries task-essential information, ESC-CR improves over complete message dropping in every reported cell while maintaining zero observed code ASR in this suite. The result is strongest for Qwen3.5-9B chain (63.00\% versus 18.50\%), showing that recovery can preserve independently verified information instead of reducing the system to message deletion. This conclusion is limited to the local trusted-evidence construction used by the benchmark.
	
	\section{Conclusion}
	
	We studied secure inter-agent code generation when messages contain both task-relevant information and unauthorized instructions. We introduced Executable Semantic Commitments with Clean-Room Recovery (ESC-CR), which separates information from authority, enforces task-scoped effects at an external release boundary, and reconstructs evidence-backed context without reusing tainted state.
	Under matched generation budgets, polluted-context retry remained ineffective, while complete message removal lost information required by communication-essential tasks. ESC-CR preserved verified claims while suppressing unauthorized releases across the evaluated code and agent settings. These results show that secure recovery is both an authorization problem and a state-management problem: rejected state must be removed without discarding independently supported task information.
	
	\clearpage
	
	\bibliography{references}
	
\end{document}